\documentclass[11pt]{article}

 \def\bb{\bibitem} \def\lb{\label}
\def\be{\begin{equation}} \def\ee{\end{equation}}
\def\ba{\begin{eqnarray}} \def\ea{\end{eqnarray}} \def\part{\partial}
\def\ol{\overline} \def\e{{\rm e}} \def\k{\kappa}
\def\X{{\bf X}} \def\P{{\bf P}} \def\L{{\bf L}} \def\J{{\bf J}} \def\S{{\bf S}}
\def\M{{\cal M}} \def\dM{\part{\cal M}} \def\R{{\cal R}} 
\def\K{{\cal K}} \def\T{{\it \Theta}} \def\g{{\cal G}} \def\l{\ol{l}}

\begin{document}
\begin{titlepage}
\date{31 January 2003}

\title{
   \begin{flushright} \begin{small}
     LAPTH-962/03
  \end{small} \end{flushright}
\vspace{1.5cm} Black hole mass and angular momentum\\ in 2+1 gravity}
\author{G\'erard
Cl\'ement \thanks{Email: gclement@lapp.in2p3.fr}  \\ \\ {\small
Laboratoire de  Physique Th\'eorique LAPTH (CNRS),} \\ {\small
B.P.110, F-74941 Annecy-le-Vieux cedex, France}}
\maketitle

\begin{abstract}
We propose a new definition for the mass and angular momentum of
neutral or electrically charged black holes in 2+1 gravity with two
Killing vectors. These finite conserved quantities, associated with
the $SL(2,R)$ invariance of the reduced mechanical system, are shown
to be identical to the quasilocal conserved quantities for an improved
gravitational action corresponding to mixed boundary conditions. They
obey a general Smarr-like formula and, in all cases investigated, are
consistent with the first law of black hole thermodynamics. Our
framework is applied to the computation of the mass and angular
momentum of black hole solutions to several field-theoretical models.
\end{abstract}
\end{titlepage}
\setcounter{page}{2}

\section{Introduction}

A long-standing problem in general relativity is that of a
satisfactory definition of the total energy of a self-gravitating
distribution of matter. The modern approach to this question uses the
idea of quasilocal energy \cite{RT, Wa93, BY, Hay, HaHo96, kblb,
Chen, Booth, BLY} (a more extensive bibliography is given in the last
three references). From an action functional for the gravity-matter
system, with boundary conditions on a given hypersurface, one derives
canonically a Hamiltonian, given by the sum of a bulk integral, which
vanishes on shell, and of a surface term. The quasilocal energy is
defined as the on-shell value of the Hamiltonian. This is generically
divergent in the limit where the spatial boundary goes to infinity,
but can be made finite by substracting the contribution of an
appropriate background. From this Hamiltonian, one can also define in
the rotationally symmetric case a quasilocal angular momentum. The
various possible choices of boundary conditions (for instance
Dirichlet or Neumann boundary conditions) and of backgrounds lead to
different possible forms for these conserved quantities, which
however give the same result in the case of asymptotically flat
configurations with a Minkowski background. Investigations of
conserved quantities in non-asymptotically flat spacetimes have
concerned mainly asymptotically dS or AdS spacetimes \cite{AbDe82, AsMa84,
HeTe85,ACOTZ}. The definition of the total energy given in \cite{AbDe82} has
been shown to agree with the Dirichlet quasilocal energy
\cite{HaHo96}. The first application of the quasilocal formalism to
AdS black holes was given in \cite{BrCrMa94}. The quasilocal framework 
has also recently been applied to the computation of the mass and angular
momentum of non-asymptotically flat, non-asymptotically AdS dilatonic 
black holes \cite{newdil}.

A testing ground for these various possible definitions of conserved
quantities is provided by 2+1 gravity, as all known (2+1)-dimensional
black hole spacetimes are non-asymptotically flat. Indeed, it can be
argued that the departure from asymptotic flatness in the 2+1 case is
such that it does not make sense to impose boundary conditions at
infinity. Rather, boundary conditions should be imposed on some
(arbitrarily chosen) timelike surface. It turns out that this is
possible because of two peculiarities of general relativity in 2+1
dimensions. First, this theory is, as Newtonian gravity in four
dimensions, dynamically trivial \cite{DJH}, so that (unlike the case
of general relativity in 3+1 dimensions), the total mass and angular
momentum inside a closed one-surface (i.e.\ a closed line) can in
principle be written as some fluxes through that surface. Second,
these black hole spacetimes are stationary and rotationally  
symmetric, i.e.\
they have two commuting Killing vectors. The resulting  $SL(2,R) \sim
SO(2,1)$ invariance leads, by the usual Noether mechanism, to three
constants of the motion \cite{EML}. We shall argue, and check on a
number of specific examples, that two of these constants may be
identified with mass and angular momentum, which can therefore be
computed on a circle of arbitrary finite radius. This identification
will be supported by a quasilocal computation, our
surface-independent energy corresponding to a well-defined finite
part of the canonical quasilocal energy for Dirichlet boundary
conditions, while our angular momentum coincides with the canonical
quasilocal angular momentum.

In the next section, we recall the dimensional reduction of 2+1
Einstein gravity with two Killing vectors to a mechanical problem.
The residual $SL(2,R)$ invariance of the reduced theory leads to the
conservation of a super angular momentum, two components of which are
associated with mass and angular momentum. This association is
checked in the case of the black-hole solutions to vacuum Einstein
gravity with a negative cosmological constant. We also show, by
evaluating our surface-independent energy and angular momentum on the
horizon, that they satisfy a Smarr-like formula. We then relate in
Sect.\ 3 our conserved quantities to the same quantities computed in
the canonical quasilocal approach, and show that our energy and
angular momentum may be derived canonically from an improved action,
corresponding to mixed Dirichlet-Neumann boundary conditions. Our
framework is applied in Sect.\ 4 to the computation of the mass of
black hole solutions to two gravitating scalar field models. Sect.\ 5
is devoted to the extension of our approach to the case of matter
gauge fields, which is then tested in Sect.\ 6 on the example of
black hole solutions to Einstein-Maxwell gravity with a negative
cosmological constant.

\setcounter{equation}{0}
\section{Conserved quantities from angular momentum in minisuperspace}
We consider a $3$-dimensional
Lorentzian spacetime $\M$ with metric $g_{\mu\nu}$ (we
use the Landau-Lifshitz conventions, except for the signature of the
metric which is $- + +$) and boundary $\dM$. This boundary consists
of initial and final spacelike surfaces
(two-surfaces in the present case) $\Sigma_{t_1}$ and $\Sigma_{t_2}$,
and a timelike surface $\Sigma^{\rho}$ (not necessarily at spatial infinity),
which we shall assume to be orthogonal to the $\Sigma_{t}$. In the
canonical $1 + 2$ ADM decomposition \cite{ADM}, the metric on $M$
is written as
\be\lb{adm}
ds^2 = -N^2\,dt^2 + h_{ij}(dx^i + V^i\,dt)(dx^j + V^j\,dt),
\end{equation}
where $h_{ij}$ is the induced metric on $\Sigma_t$. The two-surfaces
$\Sigma_t$ and $\Sigma^{\rho}$ intersect on a one-surface
$S_t^{\rho}$, with induced metric $\sigma_{\mu\nu} = h_{\mu\nu} -
n_{\mu}n_{\nu}$, where $n^i$ is the unit normal to $\Sigma^{\rho  }$.
The action for general relativity with Dirichlet boundary conditions
on $\dM$ is
\be\lb{ac}
I_D = \int_{\M}\left(\frac1{2\k}\R + {\cal L}_m\right) +
\frac1{\k}\oint_{\Sigma_{t_1}}^{\Sigma_{t_2}}\K -
\frac1{\k}\oint_{\Sigma^{\rho}}\T
\ee
($\k = 8\pi G$), where $\R$ is the Ricci scalar density, $\K$ and
$\T$ are the traces of the extrinsic curvature densities of
$\Sigma_t$ and $\Sigma^{\rho}$, and ${\cal L}_m$ is the matter
Lagrangian density. We will first consider scalar matter described by
a set of scalar fields, the case of vector (gauge) fields shall be
treated in Sect.\ 5. The action (\ref{ac}) diverges for noncompact
spatial geometries, in which case one considers rather the relative
action, defined as the difference between the action evaluated on the
configuration $g, \phi$ (where $\phi$ stands for the matter fields)
and that evaluated on a background configuration $g_0, \phi_0$ (not
necessarily flat empty space),
\be
I_R = I - I_0.
\ee

We specialize to configurations with two
commuting Killing vectors, one timelike and one spacelike, and recall
the dimensional reduction procedure of \cite{EL, EML}. We may
choose adapted coordinates on ${\M}$ such
that
\be\lb{an}
ds^2 = \lambda_{ab}\,dx^a\,dx^b + \zeta^{-2}R^{-2}\,d\rho^2 \quad
(R^2 = -\det\lambda)
\ee
(where $x^0 = t$, $x^1 = \varphi$ and the various fields depend only
on the ``radial'' coordinate $x^2 = \rho$). This
ansatz breaks the original 3-dimensional diffeomorphism invariance
down to the product of $SL(2,R)$ (linear transformations in the
2-Killing vector space) with 1-dimensional diffeomorphisms
(reparametrizations of $\rho$). The local isomorphism of $SL(2,R)$
with the 3-dimensional Lorentz group $SO(2,1)$ suggests the
parametrization
\be\lb{vec}
\lambda = \pmatrix{
T + X & Y \cr
Y & T - X}
\ee
where the vector $\bf X$ spans a Minkowski space, i.e.
\be\lb{mink}
R^2 = {\bf X}^2 = \eta_{AB}X^AX^B = -T^2 + X^2 + Y^2\,.
\ee
Stationary solutions correspond to ``spacelike'' paths $\X(\rho)$ with
$R^2 > 0$. Intersections of these paths with the future light cone
($R^2 = 0$, $T > 0$) correspond to event horizons, while intersections
with the past light cone ($R^2 =0$, $T < 0$) correspond to naked
singularities \cite{EL}\footnote{Note that the sign convention for
$\X$ in Ref. \cite{EL} is opposite to that chosen here.}.
Our metric ansatz (\ref{an}) may also be written in terms of
null target space coordinates
\be
U = T + X, \quad V = T - X \quad (R^2 = Y^2 - UV),
\ee
as
\be\lb{anull}
ds^2 = U\,dt^2 + 2Y\,dt\,d\varphi + V \,d\varphi^2 +
\zeta^{-2}R^{-2}\,d\rho^2.
\ee
The canonical, $1+2$ metric ansatz (\ref{adm}) is related to this
$2+1$ ansatz by
\be\lb{tra}
N^2 = \frac{R^2}V, \;\; V^{\varphi} = \frac{Y}V, \;\; h_{\rho\rho} =
n_{\rho}^2 = \zeta^{-2}R^{-2}, \;\; h_{\varphi\varphi} =
\sigma_{\varphi\varphi} = V.
\ee

The Christoffel symbols for the metric (\ref{an}) are
\be\lb{chris}
\Gamma^a_{2b} = \frac12\chi^a_{\;b}\,, \quad
\Gamma^2_{ab} = -\frac12\zeta^2R^2(\lambda\chi)_{ab}\,, \quad
\Gamma^2_{22} = - R^{-1}R' - \zeta^{-1}{\zeta}'\,,
\ee
where the prime stands for $d/d\rho$, and
\be
\chi = \lambda^{-1}{\lambda}' = R^{-2}\pmatrix{
RR' - \ell^{Y} & - \ell^T + \ell^ X \cr
\ell^T + \ell^X & RR' + \ell^{Y}}\,,
\ee
where
\be
\ell^A = \eta^{AB}\epsilon_{BCD}X^C{X'}^D
\ee
(with $\epsilon_{012} = +1$) are the contravariant components of 
the wedge product ${\bf\ell} = \X\wedge{\X}'$. The corresponding 
Ricci scalar density is
\ba\lb{R}
\R & = & \zeta\left({R'}^2 - \frac14R^2\,{\rm Tr}(\chi^2)\right) -
(2\zeta RR')' \nonumber \\ & = & \frac12\zeta{{\X}'}^2 - (2\zeta
RR')'.
\ea
Using $n^i = \delta^i_2\,\zeta R$, we obtain for the trace of the
extrinsic curvature density of the cylinder $\Sigma^{\rho}$
\be\lb{Theta}
\T = - \sqrt{|\lambda|}\lambda^{ab}\nabla_a n_b = - \zeta RR',
\ee
while the trace of the extrinsic curvature of $\Sigma_t$ vanishes for
these stationary configurations, so that the purely
gravitational part of the action (\ref{ac}) reduces to \cite{EL}
\be\lb{lag}
I_D = \int d^2x \,\frac1{4\k}\int d\rho \,\zeta {\X'}^2\,.
\ee

We thus have reduced 2+1 Einstein gravity with 2 Killing vectors to a
``minisuperspace'' mechanical problem on the Minkowski plane. The
(repara-metrization invariant) momentum conjugate to $\X$ is
\be
\P = \frac1{2\k}\zeta{\X}'\,.
\ee
The gravitational contribution (\ref{lag}) to the action is invariant
under Lorentz transformations in target space. Assuming that the
matter scalar fields depend only on the radial coordinate $\rho$, the
matter part of the action depends only the metric component
$g_{\rho\rho}$ and on $\sqrt{|g|} = \zeta^{-1}$, and so is also
Lorentz invariant. It follows that the super angular momentum
\be\lb{orb}
\L = \X\wedge\P = \frac1{2\k}\zeta{\bf\ell}\,,
\ee
is a constant of the motion. However, only two conserved quantities
are associated with this vectorial conservation law, because of the freedom to
perform infinitesimal gauge transformations (transition to rotating frames)
$\delta\varphi = -\delta\Omega\,t$, leading to infinitesimal rotations of the
configuration vector
\be\lb{gauge}
\delta\X = \delta{\bf\Omega}\wedge\X
\ee
around the null direction $\delta{\bf\Omega} = \delta\Omega(1, 1, 0)$. As we
shall show in the next section, using the quasilocal approach, and
check on a number of specific examples in the following, the two
physical conserved quantities associated with $\L$ are the energy and
angular momentum,
\ba
E & = & - 2\pi L^Y\,, \lb{E}\\
J & = & 2\pi(L^T - L^X)\lb{J}\,.
\ea
Under an infinitesimal gauge transformation (\ref{gauge}), these
transform as
\be
\delta E = J\delta\Omega, \quad \delta J = 0.
\ee
It is easily checked that this transformation law remains exact for
finite gauge transformations. The preferred coordinate frame will as
usual be defined relative to an asymptotic corotating observer, so that
the asymptotic angular velocity
\be
\Omega_{\infty} = -\frac{Y}{V}\,\bigg|_{(\rho\to\infty)}
\ee
vanishes. The mass $M$ is obtained from the energy (\ref{E}) computed
in this frame by substracting the energy of an appropriate
background\footnote{One could also in principle substract from the
angular momentum (\ref{J}) the angular momentum of an appropriate
background. In practice, the background is usually non-rotating, so
that the angular momentum substraction constant vanishes.},
\be
M = E - E_0.
\ee

The simplest example is that of the BTZ black hole \cite{btz},
corresponding to the matter Lagrangian density ${\cal L}_m  =
-(\Lambda/\k)\sqrt{|g|}$ with a negative cosmological constant
$\Lambda = - l^{-2}$. In the ADM parametrization (\ref{adm}),
these solutions are given by
\be\lb{btz1}
N^2 = h_{rr}^{-1} = \frac{r^2}{l^2} - {\rm M} + \frac{{\rm J}^2}{4r^2}, \quad
h_{\varphi\varphi} = r^2, \quad V^{\varphi} = -\frac{{\rm J}}{2r^2}\,,
\ee
in units such that $\kappa = \pi$. This may be transformed into the
$2+1$ form (\ref{anull}) by the coordinate transformation
$r^2 = 2\rho + {\rm M}l^2/2$. The BTZ black
hole (\ref{btz1}) corresponds to the spacelike geodesic
\be\lb{btz2}
U = -2l^{-2}\rho + \frac{{\rm M}}2, \quad V = 2\rho + \frac{{\rm
M}l^2}2, \quad Y = -\frac{{\rm J}}2,
\ee
with $\zeta = 1$. Using
\ba
l^T & = & YX' - XY' \\
l^X & = & YT'  -TY' \\
l^Y & = & TX'-XT' = \frac12(VU'-UV'),
\ea
we check that the mass and angular momentum given by (\ref{E}) (with the
background corresponding to the BTZ vacuum ${\rm M} = {\rm J} = 0$)
and (\ref{J}) coincide with their BTZ values,
\be
M = {\rm M}, \quad J = {\rm J}.
\ee

We conclude this section by showing that equations (\ref{E}) and
(\ref{J}) imply the validity of the Smarr-like formula \cite{Smarr}
\be\lb{smarr}
M = -E_0 + \frac12 T_H S + \Omega_h J
\ee
for any black hole configuration. In (\ref{smarr}), $T_H$ is the
Hawking temperature, defined as the inverse of the period in imaginary time,
\be\lb{temp}
T_H \equiv \frac1{2\pi}\,n^{\rho}\part_{\rho}N\,|_{(\rho = \rho_h)} =
\frac{\zeta RR'}{2\pi\sqrt{V}}\,\bigg|_{(\rho = \rho_h)}
\ee
(where $\rho_h$ is the horizon radius, defined by $R^2(\rho_h) = 0$),
$S$ is the black hole entropy \cite{Reznik},
\be\lb{entro}
S \equiv \frac{A}{4G} = \frac{4\pi^2\sqrt{V}}{\kappa}\,\bigg|_{(\rho
= \rho_h)}
\ee
($A$ being the horizon perimeter), and $\Omega_h$ is the horizon
angular velocity
\be
\Omega_h = -\frac{Y}{V}\,\bigg|_{(\rho = \rho_h)}.
\ee
The integral mass formula (\ref{smarr}) is easily proven by collecting
the above definitions and equations (\ref{E}) and (\ref{J}) for the energy and
angular momentum evaluated on the horizon. It differs from the
integral mass formula for 3+1 dimensions\cite{BCH}
\be
M = \int_{\Sigma_t}(2\tau_a^b - \tau\delta_a^b)u^a\,d\Sigma_b + 2T_H
S + 2\Omega_h J
\ee
(where $\tau_a^b$ is the matter energy-momentum, and $u^a$ is the unit
normal to $\Sigma_t$) in two respects.
First, the (2+1)-dimensional formula involves only geometric
quantities evaluated on the horizon. Second, the numerical
coefficients are different.

\setcounter{equation}{0}
\section{Quasilocal energy and angular momentum}
We now recall the standard construction of quasilocal mass and
angular momentum. We start again from the action (\ref{ac}), without
any assumptions of symmetry. Introducing the canonical momenta
$p^{ij}$ and $p$ conjugate to $h_{ij}$ and $\phi$, and rearranging
the action, one arrives in the absence of matter gauge fields (the
case of the Maxwell field shall be considered in Sect. 5) at the form
\be\lb{canac}
I_D = \int dt \left[\int_{\Sigma_t}(p^{ij}\dot{h}_{ij} + p\dot{\phi} -
N{\cal H} - V^i{\cal H}_i) - \oint_{S_t^{\rho}}(N\epsilon +
2V_i\pi^{ij}n_j)\right],
\ee
where ${\cal H}$ and ${\cal H}_i$ are the Hamiltonian and momentum
constraints, $\epsilon=\epsilon_{(g)}$, where $\epsilon_{(g)}$  is the density
\be\lb{eg}
\epsilon_{(g)} = \frac1{\k} k\sqrt{|\sigma|},
\ee
$k$ being the trace of the extrinsic curvature of $S_t^{\rho}$ in
$\Sigma_t$,
\be\lb{k}
k = - \sigma^{\mu\nu}D_{\mu}n_{\nu}
\ee
(with $D_{\mu}$ the covariant derivative on $\Sigma_t$), and the
reduced momenta $\pi^{ij} = (\sqrt{|\sigma|}/\sqrt{|h|})p^{ij}$
are related to the extrinsic curvature of $\Sigma_t$
\be\lb{Kij}
K_{ij} = - \frac1{2N}(\dot{h}_{ij} - 2D_{(i}V_{j)})
\ee
by
\be
\pi^{ij} = \frac1{2\kappa}\sqrt{|\sigma|}(Kh^{ij}-K^{ij}).
\ee
For a configuration solving the field equations, the
constraints ${\cal H}$ and ${\cal H}_i$ vanish, and the Hamiltonian
reduces to the one-surface integral
\be\lb{ham}
H_D = \oint_{S_t^{\rho}}(N\epsilon_{(g)} + 2V_i\pi^{ij}n_j).
\ee
The quasilocal energy or mass is the difference between the value of
this Hamiltonian and that for the background evaluated with the same
boundary data for the fields $\sigma$ and $N$. The quasilocal momenta
are obtained from the Hamiltonian by carrying out an infinitesimal
gauge transformation $\delta x^i = \delta\xi^i\,t$ and evaluating the
response
\be\lb{mom}
P_i = \frac{\delta H}{\delta\xi^i} = - 2\oint_{S_t^{\rho}}
\pi_{ij}n^j\,.
\ee

What are the values of these quasilocal quantities in the stationary
rotationally symmetric case considered in Sect.\ 2? Using the
relations (\ref{tra}) between the 1+2 and 2+1 metric ansatz\"e and
the definitions (\ref{k}) and (\ref{Kij}) we obtain
\be
k = -\frac12 \zeta R \frac{V'}V, \quad K_{12} =
\frac{\sqrt{V}}{2R}\frac{VY'-YV'}V,
\ee
leading to
\be
\epsilon_{(g)} = - \frac1{2\kappa}\zeta R \sqrt{V}\frac{V'}V, \quad \pi^{12} =
- \frac1{4\k}\zeta^2R\frac{VY'-YV'}V.
\ee
Evaluating the Hamiltonian (\ref{ham}) and the momentum $P_{\varphi}$
(\ref{mom}) on a circle $\rho =$ constant, we obtain the quasilocal energy
and angular momentum
\ba
E_D & = & \frac{\pi}{\k}\zeta(UV'-YY'), \lb{qlen} \\ J & = &
\frac{\pi}{\k}\zeta(VY'-YV'). \lb{qlam}
\ea
Using Eq. (\ref{Theta}), (\ref{qlen}) and (\ref{qlam}) may be rewritten as
\ba
E_D & = & -2\pi L^Y + \frac1{2\k}\oint_{\Sigma^{\rho}}\T, \\ J & = &
2\pi(L^T-L^X).
\ea

It follows that the quasilocal energy deriving from the ``Dirichlet'' action
(\ref{ac}) on a one-surface $\Sigma^{\rho}$ is the sum
of a surface-independent term, the constant (\ref{E}), and a purely
geometric surface-dependent term proportional to the mean extrinsic
curvature of this surface. In the BTZ case, this last term ($\T =
-4\rho/l^2$) diverges when the ``radius'' $\rho$ is taken to infinity.
As we shall see on specific examples in the next two sections, this
divergence is generic. However it is well known
\cite{Chen} that the quasilocal energy is not uniquely defined, but
depends on the boundary conditions. Modifying the Lagrangian density
by the addition of a total divergence will not change the field
equations, but will modify the boundary conditions and the value of
the quasi-local energy-momentum. In the present case, the ``improved''
action which leads to the constant quasilocal energy and angular
momentum (\ref{E}) and (\ref{J}) is
\ba\lb{imp}
I & = & I_D - \frac1{4\k}\int_{\M}\part_{\mu}(\sqrt{|g|}k^{\mu}) \nonumber \\
& = & \int_{\M}\left(\frac1{2\k}\R + {\cal L}_m\right) +
\frac1{2\k}\oint_{\Sigma_{t_1}}^{\Sigma_{t_2}}\K -
\frac1{2\k}\oint_{\Sigma^{\rho}}\T,
\ea
where
\be
\sqrt{|g|}k^{\mu} = \sqrt{|g|}\left(g^{\mu\nu}\Gamma^{\lambda}_{\nu\lambda}
- g^{\nu\lambda}\Gamma^{\mu}_{\nu\lambda}\right) =
\frac1{\sqrt{|g|}}\part_{\nu}\bigg(|g|g^{\mu\nu}\bigg).
\ee
Remembering that under a variation of the boundary data on $\dM$ for
classical solutions the variation of the Dirichlet
action (\ref{ac}) is
\be
\delta I_D = \frac1{2\k}\oint_{\dM}
n_{\mu}\bigg(\Gamma^{\lambda}_{\nu\lambda}
\delta\g^{\mu\nu} - \Gamma^{\mu}_{\nu\lambda}\delta\g^{\nu\lambda}\bigg),
\ee
where $\g^{\mu\nu} = \sqrt{|g|}g^{\mu\nu}$ is the contravariant metric
density, we find that the variation of the improved action (\ref{imp})
\be
\delta I = \frac1{4\k}\oint_{\dM}
n_{\mu}\left[\bigg(\Gamma^{\lambda}_{\nu\lambda}
\delta\g^{\mu\nu} - \Gamma^{\mu}_{\nu\lambda}\delta\g^{\nu\lambda}\bigg) -
\bigg(\g^{\mu\nu}\delta\Gamma^{\lambda}_{\nu\lambda}
- \g^{\nu\lambda}\delta\Gamma^{\mu}_{\nu\lambda}\bigg)\right]
\ee
vanishes for mixed Dirichlet-Neumann boundary conditions, such as
those discussed in \cite{AnTu02}.

\setcounter{equation}{0}
\section{Einstein-scalar black holes}

In this section we apply equations (\ref{E}) and (\ref{J}) to the
computation of the mass and angular momentum of black hole solutions
to some specific gravitating scalar field models. Let us note that it
follows from the Minkowski space identity
\be
\L^2 = \frac{\zeta^2}{4\kappa^2}\bigg(-R^2\P^2+(\X\cdot\P)^2\bigg)
\ee
evaluated on the horizon $R^2 = 0$ that the constant vector $\L$ is
spacelike. Therefore we can always carry out a local
coordinate transformation in (\ref{an}) (corresponding to a
pseudo-rotation in target space) such that $L^T = L^X = 0$, i.e. $Y =
0$. This means that without loss of generality we can restrict
ourselves to static black holes. Rotating black holes can be generated
from these by the local coordinate transformation \cite{DJH}
$t = \tilde{t} + \omega\tilde{\varphi}$, $\varphi = \tilde{\varphi}$,
$\rho = \tilde{\rho}$ ($\omega$ constant), leading to the transformed
metric components
\be
\tilde{U} = U, \quad \tilde{Y} = \omega U, \quad \tilde{V} = V +
\omega^2 U,
\ee
and to the transformed energy and angular momentum $\tilde{E} = E$,
$\tilde{J} = -2\omega E$.

As our first example we consider the HMTZ black holes
\cite{hmtz}. The matter Lagrangian density is
\be
{\cal L}_m = \frac1{\pi}\sqrt{|g|}\left[-\frac12(\nabla\phi)^2 +
\frac1{8l^2}(\cosh^6\phi + \nu\sinh^6\phi)\right],
\ee
where $l$ and $\nu$ are coupling constants. The static black hole solutions
with regular scalar field given in \cite{hmtz} are, for $\nu > -1$,
\ba
ds^2 & = & - \bigg(\frac{H}{H+B}\bigg)^2F\,dt^2 +
\bigg(\frac{H+B}{H+2B}\bigg)^2\frac{dr^2}{F} + r^2\,d\varphi^2,
\lb{mHMTZ} \\ \phi & = & {\hbox{\rm arctanh}}\sqrt{\frac{B}{H+B}},
\ea
where
\be
H(r) = \frac12\bigg(r + \sqrt{r^2+4Br}\bigg), \quad F(r) =
\frac{H^2}{l^2} - (1+\nu)\bigg(\frac{3B^2}{l^2} +
\frac{2B^3}{l^2H}\bigg),
\ee
with $B$ a non-negative integration constant, and $G = 1$ ($\kappa =
8\pi$).
This is of the form (\ref{anull}) with $\rho = r$ and
\be
\zeta = \frac{H+2B}{rH}.
\ee
Using the identity
\be
H^2 = r(H+B),
\ee
we obtain
\be
\frac{U}{V} = -\frac{F}{H^2}.
\ee
Differentiating this last relation and using
\be
H' = \frac{H(H+B)}{r(H+2B)},
\ee
we obtain the mass
\be\lb{hmtzmass}
M = \frac{3(1+\nu)B^2}{8l^2}
\ee
(with the BTZ vacuum solution $B =0$ as background). This value,
which can be checked to agree with the first law of black hole
thermodynamics
\be\lb{first}
dM = T_H\,dS + \Omega_h\,dJ,
\ee
is identical to that obtained in \cite{hmtz} by a totally different
approach, namely the computation of the total charge, which involves
the knowledge of the asymptotic behaviors of the metric functions {\em
and} of the scalar field, and yields a finite result only after
substraction of the background contribution. Computation of the
Dirichlet quasilocal energy (\ref{qlen}) would also lead to a divergent
result (before background substraction), as the metric (\ref{mHMTZ}) is
asymptotically AdS.

Our second example will be the cold black hole solutions to the model
of a gravitating massless scalar field with a {\em negative}
gravitational constant $\kappa$ \cite{sig} (because $2+1$ gravity is
dynamically trivial, both signs of the gravitational constant are
allowed \cite{DJH}). The matter Lagrangian density is
\be
{\cal L}_m = -\frac12\sqrt{|g|}(\nabla\phi)^2.
\ee
The static rotationally symmetric solutions \cite{BBL, sig}
\be\lb{cold1}
ds^2 = -x^2\,dt^2 + b^2x^{2\alpha}(dx^2 + d\varphi^2), \quad \phi =
a\ln x
\ee
($\alpha = \kappa a^2/2$ and $b$ integration constants) have a Killing
horizon at $x = 0$. We shall show that the metric (\ref{cold1}) can be
extended across this horizon for a discrete set of values $\alpha < 0$
(implying $\kappa < 0$). We transform to the conformal gauge
\be\lb{cold2}
ds^2 = \bigg(\frac{|\alpha|}{b}r\bigg)^{2/\alpha}(-dt^2 + dr^2) +
\alpha^2r^2d\varphi^2
\ee
($r = (b/|\alpha|)x^{\alpha}$), and define for $\alpha \neq 0$
Kruskal-like null coordinates $\bar{u}$ and $\bar{v}$ by
\be
t+r = u = \bar{u}^{1-n}, \quad -t+r = -v = (-\bar{v})^{1-n},
\ee
with
\be
n = \frac2{\alpha+2}.
\ee
The metric (\ref{cold2}) may be written in mixed coordinates $(\bar{u},
v)$ as
\be
ds^2 = (n-1)\bigg|\frac{n-1}{bn}\bigg|^{\frac{n}{1-n}}
(1-v\bar{u}^{n-1})^{\frac{n}{1-n}}d\bar{u}dv
+\bigg(\frac{n-1}{n}\bigg)^2\bar{u}^{2(1-n)}(1-v\bar{u}^{n-1})^2d\varphi^2.
\ee
For $n > 1$ ($-2<\alpha<0$), this may be extended across the future
horizon $\bar{u} = 0$ provided $n$ is a positive integer. Extension
across the past horizon $\bar{v} = 0$ is likewise achieved under the
same condition. So there is a discrete sequence of regular black hole
solutions (\ref{cold2}) with $\alpha = 2(1-n)/n$, $n$ integer
\cite{sig}. These black holes have infinite horizon area (infinite
entropy) and vanishing surface gravity (vanishing temperature), so
that their mass cannot be computed as usual from the first law of
black-hole thermodynamics (\ref{first}).

To compute this mass, we note that the metric (\ref{cold1}) is of the
form (\ref{anull}) with $\rho = x$ and
\be
U = -x^2, \quad V = b^2x^{2\alpha}, \quad \zeta = b^{-2}x^{-1-2\alpha}.
\ee
The computation of the energy (\ref{E}) gives straightforwardly
\be
E = \frac{\pi}{\kappa}(1-\alpha).
\ee
The natural background here is the vacuum solution, (\ref{cold1}) with
$\alpha = 0$ ($n = 1$), corresponding to the (flat) rotationally symmetric
Rindler metric. So the black hole masses $M_n = E_n - E_1$ are
\be\lb{sigmass}
M_n = -\frac{\pi}{\kappa}\alpha =
\frac{2\pi}{|\kappa|}\bigg(\frac1{n}-1\bigg).
\ee
Note that while these masses are negative, they are bounded from
below, $M_n > -2\pi/|\kappa|$.

In this very special case, the difference $-(\pi/2\kappa)\zeta(R^2)'$
between the Dirichlet quasilocal energy (\ref{qlen}) and the energy
(\ref{E}) happens to be constant and dependent on the parameter
$\alpha$, so that the quasilocal (Brown--York) energy
\be
E_D = -\frac{2\pi}{\kappa}\alpha,
\ee
would lead to a value for the mass larger by a factor two than the
correct value given by (\ref{sigmass}).

\setcounter{equation}{0}
\section{Conserved quantities in Einstein-Maxwell theory}

The approach of Sects.\ 2 and 3 must be modified in the case of
matter gauge fields. To be specific we consider the case of one
Maxwell gauge field $F_{\mu\nu} = \part_{\mu}A_{\nu} -
\part_{\nu}A_{\mu}$ , with the Lagrangian density
\be
{\cal L}_m = -\frac14\sqrt{|g|}F^{\mu\nu}F_{\mu\nu}
\ee
(plus possible scalar field contributions), more general cases (such
as that of a Maxwell-Chern-Simons gauge field \cite{SD}) can be
treated similarly. In this case, the stationary rotationally symmetric
ansatz (\ref{an}) together with the ansatz for the potential $A_{\mu}$,
\be
A_{\mu}dx^{\mu} = \psi_a(\rho)\,dx^a,
\ee
reduces the gravitational plus gauge part of the action to \cite{SD}
\be\lb{lag1}
I_D = \int d^2x\,\frac12\int d\rho\,\zeta\bigg(\frac1{2\kappa}\X'^2 +
\ol{\psi}'{\bf\Sigma}\cdot\X\psi'\bigg),
\ee
where the real Dirac-like matrices
are defined by
\be
\Sigma ^0 = \left(
\begin{array}{cc}
0 & 1 \\
-1 & 0
\end{array}
\right) \, , \,\,\,
\Sigma ^1 = \left(
\begin{array}{cc}
0 & -1 \\
-1 & 0
\end{array}
\right) \, , \,\,\,
\Sigma ^2 = \left(
\begin{array}{cc}
1 & 0 \\
0 & -1
\end{array}
\right) \, ,
\ee
$\psi$ ($\psi^t$) is the column (row) matrix of elements ($\psi_0$,
$\psi_1$), and $\ol{\psi} \equiv \psi^t \, \Sigma^0$ .

The reduced problem now involves, besides the three ``vector'' coordinates
$X^A$, two ``spinor'' coordinates $\psi_a$ which may be eliminated in
terms of their constant conjugate momenta $\Pi^a = -\zeta^{-1}F^{ra}$,
\be
\Pi = \zeta\,\ol{\psi}\,'{\bf\Sigma}\cdot\X\,.
\ee
The $SL(2,R)$ invariance of the reduced action (\ref{lag1}) again
leads to the conservation of a super angular momentum vector, which is
now the sum
\be
{\J} = {\L} + {\S} \, ,
\ee
of the ``orbital'' contribution (\ref{orb}) and of the ``spin'' contribution
\be
{\S} \equiv \frac{1}{2} \, \Pi \, {\bf\Sigma} \, \psi \, .
\ee
The natural generalisation of the assignments (\ref{E}) and (\ref{J})
is now to identify the total physical energy and angular momentum with
the relevant components of $\J$, i.e.
\ba
E & = & -2\pi J^Y, \lb{E1} \\
J & = & 2\pi(J^T - J^X)\,. \lb{J1}
\ea

To check this in the quasilocal approach, we note that the Dirichlet
action can again be written in the canonical form (\ref{canac}), where
now
\be
\epsilon = \epsilon_{(g)} + \epsilon_{(m)},
\ee
with
\be
\epsilon_{(m)} = \frac1{N}A_0\pi^r,
\ee
where $\pi^r = \sqrt{|g|}F^{r0}$. The Hamiltonian (\ref{ham}) therefore becomes
\cite{HaRo95}
\be\lb{ham1}
H_D = \oint_{S_t^{\rho}}(N\epsilon_{(g)} + A_0\pi^r + 2V_i\pi^{ij}n_j).
\ee
Noting that $\pi^r = -\Pi^0$, we derive from (\ref{ham1}) the
quasilocal energy and angular momentum
\ba
E_D & = & \frac{\pi}{\k}\zeta(UV'-YY') - 2\pi\Pi^0\psi_0, \lb{qlen1} \\
J & = & \frac{\pi}{\k}\zeta(VY'-YV') + 2\pi\Pi^0\psi_1. \lb{qlam1}
\ea
Using Eq. (\ref{Theta}), (\ref{qlen}) and (\ref{qlam}) may be rewritten as
\ba
E_D & = & -2\pi J^Y + \frac12\oint_{\Sigma^{\rho}}\bigg(\frac1{\k}\T -
\Pi\psi\bigg), \\
J & = & 2\pi(J^T-J^X).
\ea
So the quasilocal total angular momentum for a charged configuration
is indeed given by (\ref{J1}), while the quasilocal total energy is
again the sum of the surface-independent value (\ref{E1}) and of a
surface-dependent term which may in principle be discarded by suitably
modifying the boundary conditions.

\setcounter{equation}{0}
\section{Einstein-Maxwell black holes}

Let us apply our formulas (\ref{E1}) and (\ref{J1}) to the
computation of the mass and angular momentum of the charged black
holes constructed in \cite{EL, SD} (see also \cite{MTZ}). These
solutions to the Einstein-Maxwell equations with a negative
cosmological constant $\Lambda = -l^{-2}$ depend on three parameters
$Q$, $r_0$ and $\omega$, and are given by
\be \lb{CS}
ds^2 = -N^2\,dt^2 + K^2\,(d\varphi + V^{\varphi}dt)^2 + \frac{r^2}{K^2}\,
\frac{dr^2}{N^2},
\ee
with
\ba
N^2 & = & \frac{r^2}{l^2K^2}\bigg(r^2-\l^2\k Q^2\ln(r/r_0)\bigg),
\quad V^{\varphi} =
- \frac{\omega}{K^2}\k Q^2\ln(r/r_0)\,, \\
K^2 & = & r^2 + \omega^2\k Q^2\ln(r/r_0), \quad
A_{\mu}\,dx^{\mu} = Q\ln(r/r_1)\,(dt - \omega d\varphi)\,, \lb{KA}
\ea
where $\l^2 = l^2 - \omega^2$, and the gravitational constant
$\k$ is assumed to be positive. It follows from the inequality
\be
r^2-\l^2\k Q^2\ln(r/r_0) \ge
\frac{\l^2\k Q^2}{2r_0^2}\bigg(1-\ln\bigg(\frac{\l^2\k
Q^2}{2r_0^2}\bigg)\bigg)
\ee
that the condition for the existence of horizons is
\be\lb{bound}
\l^2 \ge 2\e\, r_0^2/\k Q^2.
\ee
There are then two horizons, the event horizon radius $r_h$ being
given by the largest root of $N^2$,
\be\lb{hor}
r_h^2-\l^2\k Q^2\ln(r_h/r_0) = 0,
\ee
with $r_h/r_0 \ge \e^{1/2}$.
The angular velocity of the horizon is
\be
\Omega_h = \omega/l^2.
\ee
In the expression for the gauge potentials we have allowed for an
arbitrary additive parameter $\ln(r_1)$ common to $A_t$ and
$A_{\varphi}$, which does not modify the asymptotic potentials (the
choice $r_1 = r_0$ was made in \cite{SD}). The electric and
``magnetic'' charges associated with
the conjugate momenta $\Pi^a$ are related to the black hole parameters by
\be
\Pi = Q\,(1 \quad \omega/l^2)\,.
\ee

The computation of the quasilocal mass of the charged black hole is
not straightforward. The Hamiltonian (\ref{ham1}) diverges as
$r^2\ln{r}$ so that, even after substraction of the background energy,
there remains a logarithmic divergence. In \cite{MTZ} an {\em ad hoc}
renormalization procedure was used to cancel this divergence. On the
other hand in \cite{isol}, the isolated horizon framework together
with a Hamiltonian approach based on a dreibein rather than a metric
formulation led to a finite result. As we shall now show, our super
angular momentum approach also leads to finite results, which we shall
compare with those of \cite{isol}.

We first put the solution (\ref{CS}) in the form (\ref{anull}), with
\be
U = -\frac{r^2}{l^2} +\k Q^2 L, \quad V = r^2 + \omega^2\k Q^2 L, \quad Y
= -\omega\k Q^2 L, \quad \zeta = \frac1{r},
\ee
where $L(r) = \ln(r/r_0)$. Computation of the orbital and spin super
angular momenta gives
\ba
&&L^Y = \bigg(1+\frac{\omega^2}{l^2}\bigg)\frac{Q^2}4(1-2L),
\quad S^Y = \bigg(1+\frac{\omega^2}{l^2}\bigg)\frac{Q^2}2\,\hat{L}, \\
&&L^T-L^X = -\frac{\omega Q^2}2(1-2L), \quad S^T - S^X = -\omega
Q^2\,\hat{L},
\ea
with $\hat{L}(r) = \ln(r/r_1)$, leading to
\ba
M & = & -E_0 + \bigg(1+\frac{\omega^2}{l^2}\bigg)\,\mu, \lb{MCS} \\
J & = & 2\omega\,\mu, \lb{JCS}
\ea
with
\be\lb{mu}
\mu = \pi Q^2[\ln(r_1/r_0)-1/2]\,)\,.
\ee
It is easily checked that these values for the mass and angular
momentum satisfy for all values of $r_1$ the generalized
Smarr-like formula, which replaces (\ref{smarr}) in the charged case
\be\lb{smarrQ}
M = -E_0 + \frac12 T_H S + \Omega_h J + \frac12 \Phi_h \ol{Q}\,.
\ee
In (\ref{smarrQ}), $\ol{Q} = 2\pi Q$ is the electric charge,
$\Phi_h$ is the horizon electric potential
\be\lb{phih}
\Phi_h \equiv -(A_t + \Omega_h A_{\varphi}) =
-\frac{\l^2}{l^2}Q\ln(r_h/r_1),
\ee
and the Hawking temperature and black hole entropy defined in
(\ref{temp}) and (\ref{entro}) are here given by
\be\lb{TS}
T_H = \frac{\l^3}{l^3}\,\frac{\k Q^2}{2\pi
r_h}\bigg(\ln(r_h/r_0)-1/2\bigg)\,, \quad S = \frac{l}{\l}\,\frac{4\pi^2
r_h}{\k}\,.
\ee

Our mass and angular momentum (\ref{MCS}) and (\ref{JCS}) depend on
two arbitrary parameters $r_1$ and $E_0$. As mentioned in \cite{isol},
the gauge parameter $r_1$, or equivalently the boundary
value of the electromagnetic potentials on the horizon, is constrained
by the requirement that the first law of charged black hole thermodynamics
\be\lb{firstQ}
dM= T_H\,dS + \Omega_h\,dJ  + \Phi_h\,d\ol{Q}
\ee
holds. We start by evaluating the difference
\ba
dM - \Omega_h\,dJ & = & -dE_0 + \frac{\l^2}{l^2}\,d\mu \nonumber \\
& = & -dE_0 + \frac{\l^2}{l^2}\,\pi Q^2 \bigg[2(L_1-1/2)\frac{dQ}{Q} +
\frac{dr_1}{r_1} - \frac{dr_0}{r_0} \bigg],
\ea
with $L_1 = \ln(r_1/r_0)$. The variation of (\ref{hor}) and of the
entropy (\ref{TS}) leads to
\be
\frac{dr_0}{r_0} = -2(L_h-1/2)\frac{dS}{S} + \frac{d\l}{\l} +
2L_h\frac{dQ}{Q}\,,
\ee
with $L_h = \ln(r_h/r_0)$. Finally we obtain
\be\lb{firstQ1}
dM - \Omega_h\,dJ - T_H\,dS - \Phi_h\,d\ol{Q} = -dE_0 +
\frac{\l^2}{l^2}\,\pi Q^2 \bigg(\frac{dr_1}{r_1} - \frac{dQ}{Q} -
\frac{d\l}{\l} \bigg).
\ee
We assume here
\be\lb{ground}
E_0 = 0.
\ee
It then follows that the first law is satisfied provided
\be\lb{r1}
r_1 = \e^{\alpha}\sqrt{\frac{\k}2}Q\l\,,
\ee
where $\alpha$ is some fixed constant, leading to
\be\lb{mua}
\mu = \pi Q^2\bigg[\alpha + \frac12\ln\bigg(\frac{\k Q^2\l^2}{2\e
r_0^2}\bigg)\bigg].
\ee

For extreme black holes, which correspond to the minimum of
(\ref{bound}), (\ref{mua}) reduces to $\mu_{ex} = \pi Q^2\alpha$, so
that the mass and angular momentum of extreme black holes are
proportional to the arbitrary constant $\alpha$. A natural choice is
\be\lb{a0}
\alpha = 0,
\ee
such that all extreme black holes have zero mass and angular
momentum. Extreme black holes also have vanishing Hawking temperature
and (for $\alpha = 0$) vanishing horizon electric potential, but
nevertheless are classified  by two parameters, the horizon angular
velocity $\Omega_h = \omega/l^2$, and the horizon perimeter or
entropy proportional to the electric charge, $S_{ex} = (4\pi^2
l/\sqrt{2\k})Q_{ex}$. Thus, choosing the arbitrary parameters $E_0$
and $r_1$ to have the values (\ref{ground}) and (\ref{r1}) with
$\alpha = 0$ amounts to choosing, for each one-parameter family
of black holes with given values of $Q$ and $\omega$ (or $\l$), the
corresponding extreme black hole as background. It then follows from
the inequality (\ref{bound}) that $\mu$ is positive for all
nonextreme black holes, ensuring the positivity of the mass $M$.

The neutral limit to the uncharged BTZ black holes may be performed
by fixing the horizon perimeter and angular velocity, i.e. the
parameters $r_h$ and $\omega$, and taking the electric charge $Q$ to
zero. From the relations (\ref{hor}) we see that this is possible
only if $r_0$ goes to zero so that $\pi Q^2 \ln(r_h/r_0)$ and $\mu$
converge (for any fixed value of $\alpha$ in (\ref{r1})) to a fixed
limit $\mu_0$. It follows that the neutral limit is achieved by
replacing in the various metric functions of (\ref{CS}) $\kappa
Q^2\ln(r/r_0)$ by its limiting horizon value $\kappa\mu_0/\pi$. It is
easily checked that this replacement, together with the radial
coordinate transformation $r^2 = \ol{r}^2 - \omega^2\k\mu_0/\pi$
leads, for $\k = \pi$, to the BTZ metric (\ref{btz1}). On account of
(\ref{bound}) this neutral limit does not commute with the extreme
limit $\l \to 0$ , which leads to charged, massless black holes.

In order to compare our expressions for the black hole mass and
angular momentum with those
obtained in \cite{isol} (AWD), we note that in \cite{isol}
$\kappa$ has been set to $1$, while their charge parameter is related
to ours by $Q^2_{AWD} = 2\pi Q^2$. Furthermore, the gauge choice of
\cite{isol},
\be\lb{gaisol}
A_{\mu}dx^{\mu} = Q\ln(r/\l)(dt-\omega\,d\varphi) -
Q(\omega^2/2l^2)\,dt,
\ee
differs from our choice (\ref{KA}). The transformation from (\ref{KA})
to (\ref{gaisol}) may be made in two steps: first, set $r_1 = \l$
in (\ref{KA}); second, translate the electric potential by $A_t \to
A_t -Q\omega^2/2l^2$. Going back to (\ref{firstQ1}), we see that the
first step ($r_1$ constant) is consistent with the first law only if our
assumption (\ref{ground}) for the background energy is replaced by
\be
\tilde{E}_0 = -\frac{\pi Q^2}2\,\frac{\l^2}{l^2}.
\ee
Our expression (\ref{MCS}) accordingly becomes
\be
\tilde{M} = \pi Q^2\left[\bigg(1+\frac{\omega^2}{l^2}\bigg)\ln(\l/r_0)
- \frac{\omega^2}{l^2} \right]\,.
\ee
The second step then gives (using e.g.\ (\ref{smarrQ}))
\be
M_{AWD} = \tilde{M} + \frac{\pi Q^2\omega^2}{2l^2} =
\pi Q^2\left[\bigg(1+\frac{\omega^2}{l^2}\bigg)\ln(\l/r_0)
- \frac{\omega^2}{2l^2} \right]\,,
\ee
which corresponds to the value of the energy given in Eq.\ (IV.10)
of \cite{isol}. The comparison of the angular momentum values is more
straightforward. Rewriting equation (\ref{mu}) for $\mu$ as
\be
\mu = \pi Q^2[\ln(r_h/r_0) - \ln(r_h/r_1) - 1/2],
\ee
and using (\ref{hor}), we obtain
\be
J = \omega\left[\frac{A^2}{2\pi l^2\k} - 2\pi Q^2\ln\frac{A\l}{2\pi
lr_1} - \pi Q^2\right]
\ee
($A = 2\pi r_h l/\l$ being the horizon perimeter). This is equivalent
to Eq. (V.13) of \cite{isol} for the choice $r_1 = \l$.

\section{Conclusion}

We have proposed a new definition for the mass and angular momentum
of black holes in 2+1 gravity with two Killing vectors. These are
associated with two components of the super angular momentum of the
reduced mechanical system, which are finite and independent of the
one-surface on which they are computed. We have compared our approach
to the standard quasilocal approach, and showed that our mass and
angular momentum were the quasilocal conserved quantities for an
improved action corresponding to mixed boundary conditions. We have
also shown that these quantities, together with the other black hole
parameters, obey a general Smarr-like formula and, in all cases
investigated, are consistent with the first law of black hole
thermodynamics. Finally, we have tested our new definitions on the
example of several models. In the case of gravitating scalar
field models, our values for the mass and angular momentum agree with
previous independent computations.

In the case of charged black holes, our values are also consistent
with previous computations. However the situation in this case is (as
in the case of four-dimensional non-asymptotically flat charged black
holes) not fully satisfactory. For asymptotically flat charged black
holes, the electric potential $\Phi$ goes to a constant value at
spatial infinity, and there is a natural gauge $\Phi(\infty) = 0$ in
which to compute the quasilocal energy. When the electric potential
diverges at infinity, it is only defined up to an additive constant,
which leads to an inherent one-parameter additive ambiguity in the
electrostatic energy \cite{isol}. It has been suggested in
\cite{HaRo95} that the gauge should be fixed such that the electric
potential be regular on the black hole horizon. However such a
gauge-fixing is not consistent with the first law in the case of
the (2+1)-dimensional charged black holes studied in \cite{isol} and
here.

Let us also emphasize that our new definitions apply only to the case
where the matter fields depend only on the radial coordinate. For
instance, they do not apply to the conical spacetime \cite{star}
generated by a delta-function source $\delta^2({\bf x})$ (the
quasi-local computation of the energy in this case has been given in
\cite{HaHo96}). Another example where our formalism does not apply is
the regular asymptotically conical spacetime generated by a
$\sigma$-model scalar field mapping the two-plane on the two-sphere
\cite{GC76}.

Finally, let us comment on the comparison of our approach with the
counterterm approach \cite{HS,BaKr99}. In this procedure, inspired by the
AdS/CFT correspondence, the quasilocal stress-energy of gravity is
renormalized by adding to the action a finite number of boundary
curvature invariants with coefficients fixed to ensure finiteness of
the stress tensor when the boundary is sent to infinity.
This approach was very recently extended to the
case of 2+1 gravity with a minimally coupled scalar field, leading to
a finite quasilocal energy involving a counterterm which depends
explicitly on the scalar field \cite{GMT}. The resulting value of the
HMTZ black hole mass coincides with our result (\ref{hmtzmass}). The
advantage of our procedure is that, in the neutral case, our mass and
angular momentum are defined entirely in terms of boundary data (on an
arbitrary boundary) of the metric tensor field.

\section*{Acknowledgment}

I would like to thank Dmitri Gal'tsov for numerous discussions on the
subject of quasi-local energy-momentum.

\end{document}